\documentclass[showpacs,aps,prb,groupaddress,preprint]{revtex4-1}
\usepackage{amsmath}
\usepackage{amssymb}
\usepackage{graphicx,color}
\usepackage{epstopdf}

\begin{document}

\title{Electrostatically Controlled Magnetization Rotation in Ferromagnet-Topological Insulator Planar Structures}

\author{Y. G. Semenov}
\affiliation{Department of Electrical and Computer Engineering, North Carolina State University, Raleigh, NC 27695-7911}

\author{X. Duan}
\affiliation{Department of Electrical and Computer Engineering, North Carolina State University, Raleigh, NC 27695-7911}

\author{K. W. Kim}\email{kwk@ncsu.edu}
\affiliation{Department of Electrical and Computer Engineering, North Carolina State University, Raleigh, NC 27695-7911}

\begin{abstract}
An approach to the electrostatic control of $90^{\circ}$
magnetization rotation in the hybrid structures composed of
topological insulators (TIs) and adjacent ferromagnetic insulators
(FMI) is proposed and studied. The concept is based on TI electron
energy variation with in-plane to put-of plane FMI magnetization
turn. The calculations explicitly expose the effect of free energy
variability in the form of the electrically controlled uniaxial
magnetic anisotropy, which depends on proximate exchange interaction
and TI surface electron density. Combining with inherent anisotropy,
the magnetization rotation from in-plane to out-of-plane direction
is shown to be realizable for 1.7 $\sim$ 2.7 ns under the electrical
variation of TI chemical potential in the range $\pm$ 100 meV around
Dirac point. When bias is withdrawn a small signal current can
target the out-of-plane magnetization instable state to the
desirable direction of in-plane easy axis, thus the structure can
lay the foundation for low energy nonvolatile memory prototype.
\end{abstract}

%\date{\today }
\maketitle

\bigskip

Because of the absence of the natural magnetic charges and thus the leakage
conditions, magneto-electronic automata presents a great advantage of the
magnetic devices compared with their electrical counterparts in terms of the
non-volatility. Up to date the magnetization driving force in such devices
is realized via a magnetic field of the channel current or current-induced
angular momentum transfer in magnetic tunnel junction.~\cite{Zhu2008} The
current-assisted data processing in both techniques consumes much energy,
the reduction of which is one of the major concerns for present industrial
developments and academic researches. %Plus the
%sole current controllability also limit the architectural
%compatibility regarding logic implementations.
%%%%%% Judging for the present state of MRAM, I don't fell the power consumption is a too big challenge for device integration.
%%%%%% It is actually comparable with electric RAMs.
% that is a significant challenge to integrate the magnetic devices into nano-electronic circuits.

Electric field control in addition to the current induced effects would has
been pursued with multiferroic materials. The tension variation in some FM
materials can rotate their easy axis 90$^{\circ }$ jointly with
magnetization vector $\mathbf{M}$.\cite{Maruyama09,Roy11,Liu11} In
FM/piezoelectric hybrid structures such effect evokes the new paradigm of
magnetization rotation in the electrical field. However it should be noted
that the desirable piezo-effect in such structures requires large gate
electric field, which indicates the need of searching for much stronger
magnetoelectric effects in modern magneto-electronics.

Another approach relies on the effective magnetic fields $\mathbf{H}_{ex}$
mediated by the proximate exchange interactions between surface magnetic
ions and itinerant electrons (or holes) in semiconductors,\cite{Prinz90}
graphene, \cite{SKZ07,Haugen08,Dedkov11,Mandal12} or topological insulators.\cite{Kong11,Yang11}  A distinctive feature of the $\mathbf{H}_{ex}$
consists in its dependence on electron density at the interface of the
electric channel and magnetic layers \cite{Hao91,Weisheit07} rather than on
an electric current. This property of the $\mathbf{H}_{ex}$ combines the
magnetic and electric responses inside the new "meta-materials" in a
strongly correlated manner that stems from the electrostatic and quantum
mechanical nature of the exchange interaction.

Recent studies have demonstrated the electrical control efficiency of the
proximate exchange interaction (PEI) between a magnetic layer and graphene
for different spintronic applications.~\cite%
{Weisheit07,SZK08,Yokoyama08,Zhang09,Maruyama09,Saff09,Michetti11}  As an
example, the conductivity of the graphene sandwiched between ferromagnetic
insulators (FMIs) reveals the giant magnetic resistance.\cite%
{SZK08,Michetti11} On the other hand, the magnetization of FMI can be turned
and even switched by electrical variation of the carriers density in
adjacent graphene.\cite{SZKlett}
%%%%%% I don't think this is a big challenge because multilayer structure is already widely used in MRAMs. At least it seems easy to deposite multilayers of metal.
% Note, however, that the proper functioning of memory or logic devices based on proximity with conventional semiconductors or graphene have to be designed in form of complex multilayer structure with prepared hard magnets as the reference layers that raises technological difficulties.

In present letter we explore the new prospects granted by topological
insulators (TIs) and composite structures (Fig. 1) based on TIs. Specific
spin-momentum coupling in 3D TI's surface Dirac fermions can be naturally
incorporated into magneto-electric effects of TI/FMI composite structures.
\cite{Hasan10} As a result, a lot of unusual physical phenomena have been
predicted in TIs affected by adjacent FMIs.\cite%
{Garate10,Yokoyama10,Nomura10,Kong11} Here we focus on converse TI effect on
adjacent FMI. Namely, the PEI of TI surface electrons is shown to appear in
the form of FMI effective magnetic anisotropy electrically driven without
any structure tension. Combining with inherent anisotropy and anisotropy of
the sample shape, the in-plane to out-of-plane orientation transition can be
electrically stimulated by variance of TI-mediated anisotropy. This effect
is quantitatively demonstrated for uniaxial nanomagnet deposed on the
surface of TI (e.g. Bi$_{2}$Se$_{3}$). It is remarkable that the system with
perpendicular $\mathbf{M}$ appears in neutral position with respect to $%
\mathbf{M} $ returning to either of the in-plane states along easy axis.
This instability can be used for directional targeting of $\mathbf{M}$ by a
current induced torque.\cite{Yokoyama10} Small signal current is sufficient
to tilt the magnetization to the desired direction and the following switch
will be driven by the shape and intrinsic anisotropy terms. The possible
design of nonvolatile memory and logic devices\cite{Datta10} can be realized
on basis of these magnetic units.

Qualitatively the easy axis rotation can be thought as an effect of the
electron energy reducing under exchange interaction with proximate FMI. When
the magnetization is in the in-plane direction, the PEI results in a shift
of the Dirac cones in the momentum space whereas an energy gap can be
generated and the energy dispersion becomes non-linear for the out-of-plane
orientation with $M_{z}\neq 0$.\cite{Kong11} From the thermodynamic point of
view, the case with a band gap ($M_{z}\neq 0$) is expected to be
energetically more favorable with lower electronic free energy than that
with the shifted bands ($M_{z}=0$) following a similar consideration
demonstrated in bilayer graphene.\cite{SZKlett} A departure of chemical
potential $\mu $ from the energy level of the Dirac point $\varepsilon _{D}$
suppresses this tendency because of energy increase of the conduction
electrons partially compensates valence band energy decrease while gap
opening for $\mu >\varepsilon _{D}$ and fewer active electrons for $\mu
<\varepsilon _{D}$.

To proceed the quantitative analysis let us consider a thermodynamical
potential of TI electrons interacting with FMI,
\begin{equation}
\Phi _{e}(\mu )=-T\sum_{b,\mathbf{k}}\ln \left( 1+\exp \frac{\mu
-\varepsilon _{D}-\varepsilon _{b,\mathbf{k}}}{T}\right)  \label{TP}
\end{equation}%
where the sum involves electronic bands $b$ and wave vector $\mathbf{k}$, $T$
is the temperature in energy units. We are interested in alteration of $\Phi
_{e}(\mu )$ with the Fermi level varying near the Dirac point of the surface
states. Thus the rest of electronic states in Eq. (\ref{TP}) are irrelevant
and summation in $\Phi _{e}(\mu )$ can be safely restricted by the spectrum
of surface Dirac fermions with 2D effective Hamiltonian\cite{Shan10}
%%%%%% does it mean we are not considering spin splitting? I think the logic is: 1. FMI cause the band structure to change; 2. Susceptibility is determined by the changed band structure; 3. Exchange energy is determined by the resulting susceptibility. Is there any need to explictly demonstrate this here?
\begin{equation}
H=\hbar v_{F}\left[ \mathbf{\sigma }\times \mathbf{k}\right] \widehat{%
\mathbf{z}}+Dk^{2}+\mathbf{G\sigma }  \label{H}
\end{equation}%
where $v_{F}$ is the electron Fermi velocity near Dirac point, $\mathbf{%
\sigma }$ the Pauli matrix of electron spin, $D$ is the material parameter
for quadratic term. The remaining term in Eq.~(\ref{H}) describes the energy
of an electron spin in the exchange effective fields (in units of energy) $%
\mathbf{G}=G\mathbf{m}$ of the proximate FMI, $\mathbf{m=M/}\left\vert
\mathbf{M}\right\vert $ ensures the collinearity of the effective field and
FMI magnetization.
%%%%%% I just don't understand this point. Why the collinearity is ensured in such a way? Do we need some further explanation or I'm missing something?
Introducing the angles $\theta $ between $\widehat{\mathbf{z}}$ and $\mathbf{%
m}$ and $\varphi $ between $\mathbf{k}$ and planes ($\mathbf{m}$, $\widehat{%
\mathbf{z}}$) (Fig. 1) the energy spectrum of Hamiltonian (\ref{H}) can be
expressed as
\begin{equation}
\varepsilon _{b,\mathbf{k}}=D_{0}p^{2}+b\sqrt{p^{2}+G^{2}+2Gp\sin \theta
\sin \varphi },  \label{En}
\end{equation}%
where $\mathbf{p}=\hbar v_{F}\mathbf{k}$ is the electron surface momentum in
energy unit, $D_{0}=D/\hbar ^{2}v_{F}^{2}$ and $b=\pm 1$ marks the
conduction and valence bands. Equation~(\ref{En}) clearly illustrates the band
structure reconstruction lowering the energies of electrons near the top of
valence band when $\mathbf{M}$ turns from in-plane direction ($\theta =\pi
/2 $) to vertical position ($\theta =0$). The valence band electrons ($b=-1$%
) contribute to this effect while conduction band electrons ($b=1$) may
reduce it. Thus the proximity with TI plays the role of a mediator enhancing
the uniaxial FMI anisotropy along the normal $\widehat{\mathbf{z}}$.

Following Eqs.~(\ref{En}) and (\ref{TP}),  $\Phi _{e}(\mu )$ is an even
function of $G_{x}=G\sin \theta $ that determines the dependence $\Phi
_{e}(\mu )=-K_\mathrm{eff}(\mu )\sin ^{2}\theta $ and proportionality $K_\mathrm{eff}(\mu
)\sim G^{2}$ in low-order expansion on $G$. Similar relation remains for $%
\Delta \Phi _{e}(\mu )=\Phi _{e}(\varepsilon _{D})-\Phi _{e}(\mu )$,
%%%%%% I didn't change the above equation, but I do think the notation may not be accurate. Dr. Kim used $\varepsilon_D$ and $\varepsilon_F$ in the DW paper and I adopt it here, as modified in the previous part. I think this equation may be \delta phi=phi(E_F-E_d)-phi(0). Please check this point.
which reflects the variable part of the TI mediated anisotropy energy
%%%%%% this above sentance may also need to be revised.
\begin{equation}
\Delta \Phi (\mu )=A\Delta K_\mathrm{eff}(\mu )\sin ^{2}\theta .  \label{VTP}
\end{equation}%
Here $A$ is the area of FMI/TI interface and%
\begin{equation}
\Delta K_\mathrm{eff}(\mu )=f(\mu ,T)D_{0}G^{2},  \label{Keff}
\end{equation}%
where $D_{0}$ is introduced for convenience sake. Computation of Eq.~(\ref{TP}) with Eq.~(\ref{En}) confirms the validity of the definition in Eqs.~(\ref{VTP}) and (\ref{Keff}).  Function $f(\mu ,T)$ can be
numerically found in terms of Eqs.~(\ref{TP}) and (\ref{En}); Fig.~2 depicts
the results of calculations. The results indicates the amplitude of
anisotropy changes with surprisingly minor temperature effect but
significantly asymmetry in chemical potential variation over conduction band
($\mu -\varepsilon _{D}>0$) and valence band ($\mu -\varepsilon _{D}<0$).
%%%%%% I suggest we use the notation as in the final DW paper to avoid confusion. Let's use E_F for Fermi level and E_D for the level of the Dirac point. I defined \mu as E_F-E_D previously. If this is again confusing, we may get rid of \mu completely, but just using E_F, E_D. I think that will be clear.
Consequently, Eq.~(\ref{Keff}) predicts the variation range of
anisotropy as large as 0.5 meV/nm$^{2}$ provided the strength of exchange
interaction is $G=40$ meV.\cite{Kong11} In such a case a contact area
section of tens nanometers may supply the anisotropy energy change of
eV-scale that guarantees the nonvolatile magnetic bi-stability at room
temperature.

Whether the variation of $\Delta K_\mathrm{eff}(\mu )$ is capable to cant over $%
\mathbf{M}$ depends on symmetry and magnitude of total magnetic energy $F$
of the nanomagnet. It is convenient to approximate the actual shape of the
FMI (e.g. rectangular parallelepiped) by triaxial ellipsoid with homogeneous
magnetization over the whole volume $V$.
%%%%%% I think the scale factor should be mentioned here. This factor changes the barrier from twenty some kBT to about 50kBT.
%%%%%% Plus, in the calculation, I see that this factor is also applied to induced anisotropy and current eff field. But I think the ellipsoidal approx. does't affect these two and the scale factor should not be applied.
At such approach the free energy can be simplified to normal form
\begin{equation}
F=2\pi V\sum_{i,j}N_{ij}M_{i}M_{j}+U_{an}(\mathbf{m}),  \label{Free}
\end{equation}%
where first term describes the energy of demagnetizing field, and last one
represents the magnetic anisotropy, which includes independent on $\mu $
part $\Phi _{e}(\varepsilon _{D})$ of PEI and tension-mediated term with
uniaxial anisotropy, $N_{ij}$ the components of demagnetizing tensor. In the
principal coordinates $i,j=x,y,z$, this tensor becomes diagonal, $%
N_{ij}=N_{i}\delta _{i,j}$, with principal values
\begin{equation}
N_{i}=\frac{a_{x}a_{y}a_{z}}{2}\int_{0}^{\infty }\frac{dq}{Q_{i}\sqrt{%
Q_{x}Q_{y}Q_{z}}}  \label{DMT}
\end{equation}%
where $\sum_{i}N_{i}=1$ and $Q_{i}=q+a_{i}^{2}$ with lengths $%
a_{x},a_{y},a_{z}$\ of the ellipsoid principal axes. If they coincide with
crystalline axes, the free energy per unit volume can be brought into
canonical form
\begin{equation}
\frac{F}{V}=2\pi M^{2}\sum_{i=x,y,z}N_{i}m_{i}^{2}+u_{an}(\mathbf{m})
\label{FE}
\end{equation}%
with $u_{an}(\mathbf{m})=U_{an}(\mathbf{m})/V$. In the case of planar FMI/TI
structure (i.e. $N_{z}>>N_{x},N_{y}$) the demagnetizing field tends to
establish the hard axis along normal $\widehat{\mathbf{z}}$ to interface.
So, in soft magnetic materials like Yttrium Iron Garnet, the proper choice
of FMI sizes can ensure the electrical switch between in-plane to out of
plane $\mathbf{M}$ direction. This condition, however, is not enough to
guarantee the nonvolatile bistability because there is a path that connects
the states $M_{x}$ and $-M_{x}$ through the saddle points $\pm M_{y}$.
Apparently demagnetizing field can also establish the in-plane anisotropy in
the structures with shorter FMI layer length in $y$-direction, but desirable
retention time with saddle point height $\sim 2\pi M^{2}(N_{y}-N_{x})V=$ 1
eV can be reached at least in sub-micron device sizes (200 -300 nm) as $%
N_{y} $ is still relatively small compared with $N_{z}$.

An example of the alternative approach assumes the FMI with intrinsic
uniaxial crystalline anisotropy in form of hard $y$-axis. In such a case the
anisotropy term is
\begin{equation}
u_{an}(\mathbf{m})=K_{1}m_{y}^{2}+[K_{u}+\frac{1}{t}\Delta K_\mathrm{eff}(\mu
)](1-m_{z}^{2}),  \label{MA}
\end{equation}
%%%%%% As we've discussed in the Email that K1 should be around 0.05 meV/nm^3, otherwise the energy rise to over 100 kBT abruptly which makes the figure ugly. Should we mention the value here and is it still typical?
where $K_{1}>0$, $K_{u}$ accumulates the effects of crystal lattice
distortion, which, in turn, vary with crystal doping, the strain at
interfaces with TI and spacer layer attached on opposite sides, and PEI
accounted for $\Phi _{e}(\varepsilon _{D})$. Thus, this parameter depends on
particular device implementation.

The total magnetic energy expressed in spherical coordinate of $\mathbf{m}$
(i.e., $m_{x}=\sin \theta \cos \varphi ;m_{y}=\sin \theta \sin \varphi
;m_{z}=\cos \theta $, see Fig.~1) is shown in Fig.~3(a) and 3(b) for two
chemical potentials. The graph clearly illustrates the alternation of energy
valleys and energy hills located along in-plane directions, $\theta =\pi /2$
and $\varphi =0$ or $\pi $, [Fig. 3(a)] and in the normal directions, $%
\theta =0$ or $\pi $, [Fig. 3(b)], as chemical potential changes. A small
perturbation of unstable states at tops of energy hills may guide the
following relaxation into desirable valley. For such perturbation we suggest
to evoke a proximity induced effective magnetic field $\mathbf{B}^\mathrm{eff}=%
\left[ \mathbf{J}\times \widehat{\mathbf{z}}\right] G/ev_{F}L_{z}M$ ($e$ is
the electron charge, $L_{z}$ is the FMI thickness) which accompany the spin
polarized TI surface current $\mathbf{J}$.\cite{Yokoyama11,DasSarma10} This
field can be used to choose the path to the desired valley by the current
direction. Figure~3(c) and (d) illustrates the distortion of energy relief under
small electric current. All together Fig.~3(c)-(d) exhibit the
possibility of the electric field control of magnetization turn and even
switch provided appropriate handling by bias voltage and signal current.

The dynamics of magnetization turn determines the switching speed. A common
approach to this problem is based on the solution $\mathbf{m=m}(t)$ of the
Landau-Lifshitz-Gilbert equation, which, in the case of mono-domain
approximation with homogeneous magnetization, takes the form:
\begin{equation}
\frac{\partial \mathbf{m}}{\partial t}=-\gamma \mathbf{m}\times \mathbf{H}%
_{eff}+\alpha \mathbf{m}\times \frac{\partial \mathbf{m}}{\partial t}.
\label{LLG}
\end{equation}%
where $\gamma $ is the gyromagnetic ratio, $\alpha $ the Gilbert damping
factor and the effective field $\mathbf{H}_\mathrm{eff}=(\mu
_{0}M_{0}V)^{-1}\partial F/\partial \mathbf{m}$ stems from the total free
energies $F$ [Eq.~(\ref{Free})]. Note that $\mathbf{H}_\mathrm{eff}$ crucially
depends on the magnetization.\cite{Nikonov2010} For example, if the initial
magnetization $\mathbf{m}(0)$ perfectly aligned with $x$ axis and ignoring
fluctuation, the $\mathbf{H}_\mathrm{eff}$ and $\mathbf{m}$ are collinear and the system comes into frustrating (metastable) state with respect to subsequent
relaxation to the $z$ or $-z$ direction. However any thermal dispersion of $%
\mathbf{m}(0)$ may work as a start point for magnetization switch. To show
this effect we consider evolution $\mathbf{m}(t)$ starting from $\theta $
and $\varphi $ that deviate from idealistic case $\theta =90^{\circ }$ and $%
\phi =0^{\circ }$ on only $0.1^{\circ }$. The resulting switch is shown in
Fig.~4. The switch is gradually accelerated till $t=1\ ns$, after which the $%
\mathbf{H}_\mathrm{eff}$ gathers strength and steep switch occurs with oscillation.
The actual switch time is about 1.7 ns marked by $m_{z}$ changing from 0.1
(point B) to 0.9 (point A), while it may take around 1 ns for the switch to
be initiated.

To see the dispersion of switch time, we initializes the random array of
magnetization states according to Boltzmann distribution at room
temperature. The dispersion of the shift is marked by the square point A
(when $m_{z}\geqslant 0.9$) in Fig.4 and the range is indicated by the
corresponding arrow. This means the total $90^{\circ }$ switch time varies
from $1.7$ ns to $2.7$ ns because of thermal fluctuation. One possible
solution to the undetermined switch time is to use a TI surface current to
initiate the switch by current induced torque. Thus $180^{\circ }$ switches
can be achieved.

In conclusion, a magneto-electric effect in form of electrically controlled
magnetic anisotropy has been presented for the structure taking advantage of
the proximity exchange interaction between a TI and a FMI layer. The effect
is based on exchange energy correlation with weak temperature dependence and
it is more respondent for the change in valence band than that in the
conduction band. This interaction effectively induces a change of the
perpendicular anisotropy energy of the FMI layer which results in a
thermally initiated magnetization rotation from in-plane to out-of-plane.
The initiating time can be as long as $1.4~ns$ if well aligned and the
switch time is about $1.7~ns$. A small signal current can be applied to both
initiating the switch and to determine the magnetization after withdrawing
the bias by tilting the magnetization to the desired direction at the
beginning of the relaxation.

This work was supported, in part, by the SRC Focus Center on Functional
Engineered Nano Architectonics (FENA) and the US Army Research Office.

\newpage

\begin{center}
\begin{figure}[tbp]
\includegraphics[viewport=120 340 450 500,scale=0.6,angle=0]{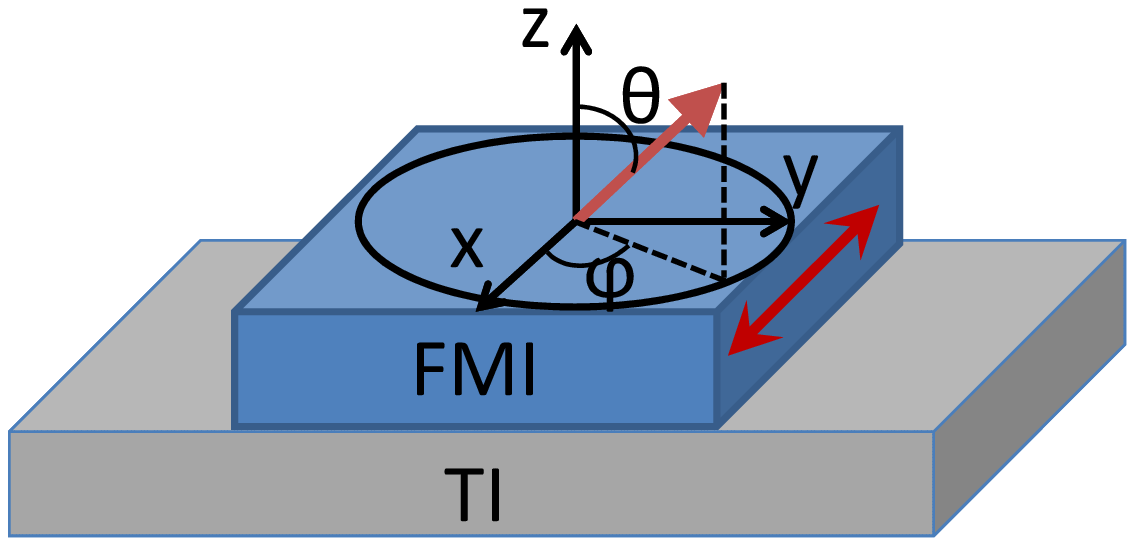}
\caption{(Color online) Schematic illustration of the structure composed of
ferromagnetic insulator layer (FMI) deposed on topological insulator (TI).
The controlling gate is not shown. The $x,y,z$ are the reference frame used
in the paper. The easy axis (two-sided arrow) is along x-direction without
gate voltage. The $\protect\theta $ and $\protect\varphi $ are the spherical
coordinates of the off-balance magnetization vector $\mathbf{M}$. A signal
current can be applied to the TI surface to tilt the magnetization.}
\end{figure}
\end{center}

\newpage

\begin{center}
\begin{figure}[tbp]
\includegraphics[scale=.8,angle=0]{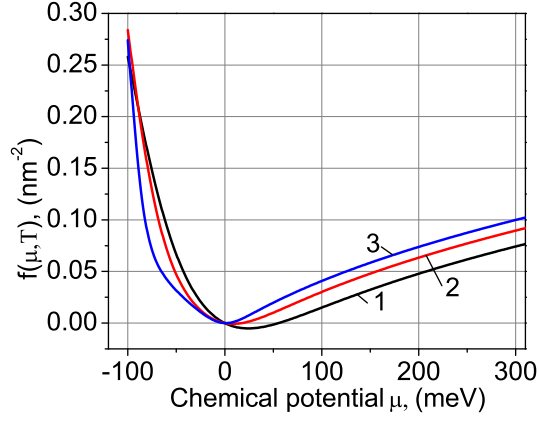}
\caption{(Color online) Parameter $f(\protect\mu ,T)$ reflecting the TI
mediated anisotropy as a function of chemical potential $\protect\mu $ at
three temperatures $300$ K (curve 1), $200$ K (curve 2) and $77$ K (curve
3). Calculations use the parameters of Bi$_{2}$Se$_{3}$ for Hamiltonian (%
\protect\ref{H}): $v_{F}=6\cdot 10^{7}$ cm/s; $D=13.0$ eV$\cdot \mathring{A}%
^{2}$.}
\end{figure}
\end{center}

\newpage

\begin{center}
\begin{figure}[tbp]
\includegraphics[scale=.6,angle=0]{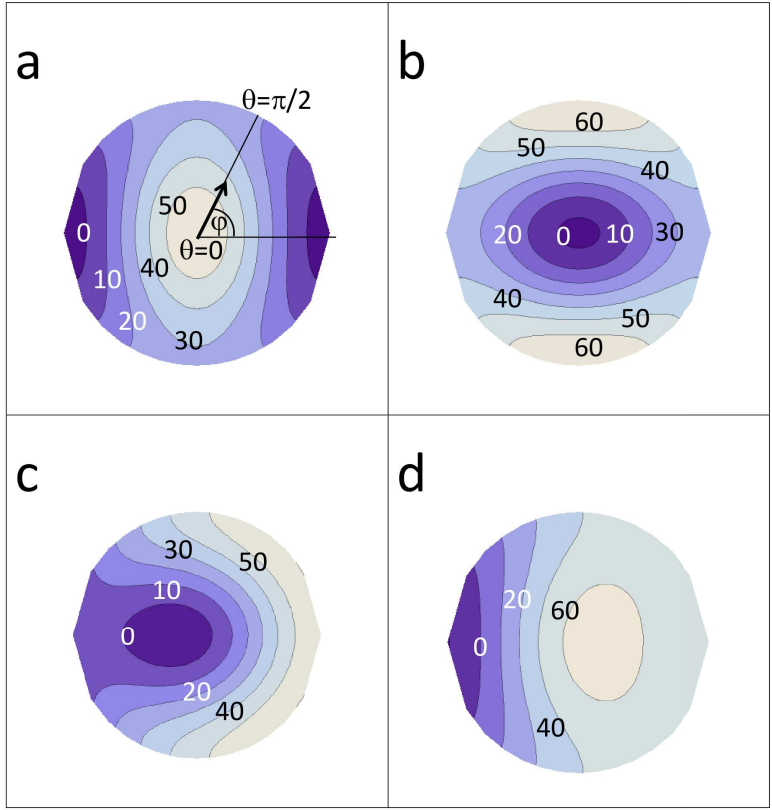}
\caption{(Color online). Energy landscape in polar coordinates $\theta $ (radial variable) and $ \varphi $ (azimuthal variable, Fig. 1) of magnetization direction. The equal-energy curves are calculated for sizes of FMI planar layer 50$\times $
50$\times $2 nm$^{3}$, $M=4\pi \times 160$~G, $K_{1}=K_{u}=0.05$
meV/nm$^{3}$ (a) Chemical potential and signal current are $\mu
-\varepsilon _{D}=-90$ meV and $J=0$. Two symmetrical valleys
correspond to $\mathbf{M}$ orientations along easy axis $x$. (b)
Easy axis is turned to z-direction with $\mu -\varepsilon _{D}=0$
and $J=0$. (c) Signal current shifts the valleys and tilts
magnetization to $-x$ direction($\mu -\varepsilon _{D}=0$ meV and
$J=1$ $\mu $A/nm) that drives the $\mathbf{M}$ relaxation to the
reversal
direction pointed at figure (d) with ($\mu -\varepsilon _{D}=-90$ meV and $
J=1$ $\mu $A/nm).}
\end{figure}
\end{center}

\newpage

\begin{center}
\begin{figure}[tbp]
\includegraphics[scale=.3,angle=0]{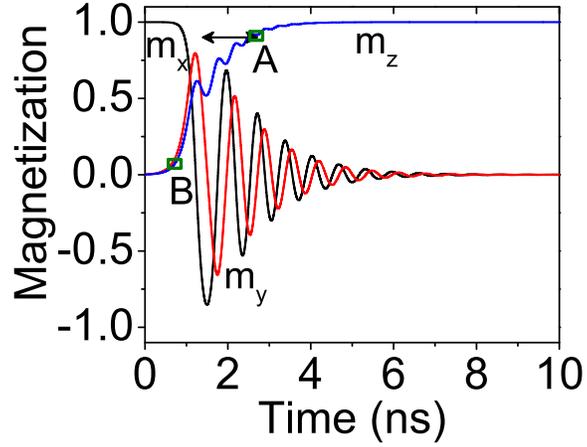}
\caption{(Color online) Anisotropy driven switching trajectory. The
initial magnetization is set to $\protect\theta=89.9^{\circ}$, $\protect\phi%
=0.1^{\circ}$ to enable the initiation of the switch. Square B ($m_{z}=0.1$)
and square A ($m_{z}=0.9$) mark the starting and ending of the switch
respectively. The switching time is around $1.7~ns$ while the initiating
time can be as long as $1.4~ns$ due to thermal fluctuation at 300 K.The
finishing point A thus varies in the range indicated by the double arrow.}
\label{switch}
\end{figure}
\end{center}

\end{document}